\documentclass[prd,amsmath,aps,floats,twocolumn,amssymb, floatfix, superscriptaddress, nofootinbib, preprintnumbers]{revtex4}
\usepackage{graphicx}
\usepackage{wasysym}
\usepackage{amsfonts}
\usepackage{subfigure}
\usepackage{hyperref}
\usepackage{color}        
\usepackage{ulem}
\usepackage{aas_macros}

\newcommand{\be}{\begin{equation}}
\newcommand{\ee}{\end{equation}}
\newcommand\xit{x^t}
\newcommand\ct{C^t}
\newcommand\cinv{C^{-1}}
\newcommand\ctinv{(C^{t})^{-1}}
\def\eea{\end{eqnarray}}
\newcommand{\ec}[1]{Eq.~(\ref{eq:#1})}
\newcommand{\Ec}[1]{(\ref{eq:#1})}
\newcommand{\eql}[1]{\label{eq:#1}}
\def\bea{\begin{eqnarray}}
\def\eea{\end{eqnarray}}
\def\vs{\nonumber\\}

\begin{document}

\preprint{LLNL-JRNL-632261}

% \title{covariance estimator error}         % Enter your title between curly braces
\title{The Effect of Covariance Estimator Error on Cosmological Parameter Constraints}

\author{Scott Dodelson}
\affiliation{Fermilab Center for Particle Astrophysics, Fermi National Accelerator Laboratory, Batavia, Illinois 60510-0500}
\affiliation{Kavli Institute for Cosmological Physics,  Enrico Fermi Institute, University of Chicago, Chicago, Illinois 60637}
\affiliation{Department of Astronomy \& Astrophysics, University of Chicago, Chicago Illinois 60637}

\author{Michael D. Schneider}        % Enter your name between curly braces
\affiliation{Lawrence Livermore National Laboratory, P.O. Box 808 L-210, Livermore, CA 94551}
\affiliation{Department of Physics, University of California, Davis, One
Shields Avenue, Davis, CA 95616}
%\date{\today}
\begin{abstract}
Extracting parameter constraints from cosmological observations requires accurate determination of the covariance matrix for use in the likelihood
function. We show here that uncertainties in the elements of the covariance matrix propagate directly to increased uncertainties
in cosmological parameters. When the covariance matrix is determined by simulations, the resulting
variance of the each parameter increases by a factor of order $1+N_b/N_s$ where $N_b$ is the number of bands in the measurement and $N_s$ is the number of simulations. 
\end{abstract}

\maketitle
\newcommand\data{x^{\rm d}}

\section{Introduction}

Upcoming galaxy surveys~\cite{Annis:2005ba,Schelgel:2011zz,Abdalla:2012fw,Abate:2012za,amendola12} aim to measure cosmological parameters at the percent level. Achieving this lofty goal will require overcoming
a number of well-known theoretical systematics: bias in translating the matter distribution to the galaxy distribution~\cite{Mehta:2011xf,vandenBosch:2012nq}, 
uncertainties in the predictions for the dark matter spectrum~\cite{Huterer:2004tr,Lawrence:2009uk},
baryonic contamination of the power spectrum in weak lensing~\cite{Rudd:2007zx,Zentner:2012mv}, outliers in photometric redshifts~\cite{Hearin:2010jr}, accurate predictions of the halo mass function~\cite{Cunha:2009rx},
and many others. 

All of these are tied to making accurate predictions for the cosmological observable, be it cluster abundance, weak lensing power spectrum, or
the position of the Baryonic Acoustic Oscillation peaks. Here we focus on the effect of uncertainty not in the observable but in the covariance matrix
of the observable, an essential ingredient in transforming the predictions and observations into parameter constraints. For simplicity throughout, we focus on the case when the likelihood is Gaussian so parameter constraints are obtained by minimizing
\be\label{eq:chisq}
\chi^2(p) = \sum_{i,j=1}^{N_b} \left( \data_i - x_i(p)\right)C^{-1}_{ij}  \left( \data_j - x_j(p)\right) 
\ee
where $p$ is the set of parameters; $\data_i$ is the data collected in $N_b$ bands (for example, the power spectrum of weak lensing at various multipole moments and redshifts or the cluster abundance in mass and redshift bins); $x_i(p)$
is the set of predictions for these measurements which depend on the parameters; and $C$ is the covariance matrix. We assume here that $C$ is independent of $p$ and therefore do not include the $\ln|C|$ normalization term in Eq.~(\ref{eq:chisq}).

In this language, most of the work about systematics to date has been directed at obtaining accurate predictions for the $x_i(p)$, while here we focus on the effect of mis-estimating the covariance matrix $C$. 
Previous work on covariance errors focused on the bias in the inverse covariance estimate~\cite{hartlap07} and uncertainties in parameter errors~\cite{Taylor:2012kz}.
Specifically, Ref.~\cite{hartlap07} showed that a statistical error in the covariance matrix estimator leads to a
multiplicative bias in the inverse covariance, or precision, matrix. This bias can be easily corrected with by multiplying the 
precision matrix estimator with a known factor depending on the number of samples used to estimate the sample covariance.
Ref.~\cite{Taylor:2012kz} identified a separate uncertainty in the covariance matrix (and precision matrix) similar to the 
result derived here, but expanded only to linear order in the sample covariance error. At linear order, Ref.~\cite{Taylor:2012kz}
showed that the inferred model parameter cosntraints cannot be known precisely due to the error in the covariance. 
Ref.~\cite{hamimeche09} also derive a result similar to ours (their Appendix~A), 
but again including only the first term in a Taylor expansion in the covariance estimator error. Ref.~\cite{hamimeche09} differ 
from Ref.~\cite{Taylor:2012kz} in finding an increase in the inferred parameter errors in addition to uncertainty in those 
errors. But Ref.~\cite{hamimeche09} disagrees with this work in the size of the increase in the parameter errors.

Here we derive an expression for the additional variance of estimators of parameters due to the uncertainties in the covariance matrix, 
expanded consistently to quadratic order in the precision matrix error. Unlike Ref.~\cite{Taylor:2012kz}, but similar to Ref.~\cite{hamimeche09} the higher order error term we consider leads to an increase in the inferred parameter errors.
We then focus on the case when the covariance matrix is estimated from simulations and dub the additional uncertainty {\it covariance estimator error}. Covariance estimator error is straightforward to compute when the measurements $x$ are Gaussian distributed, the dependence of the covariance on cosmology is neglected, and the sample covariance estimator is used. Then, the covariance estimator error enhances the variance of every parameter by a factor of order $(1+N_b/N_s)$ with $N_s$ the number of simulations used for the estimate. We go beyond the Gaussian case with the example of the weak lensing power spectrum, where we use existing simulations to compute the covariance estimator error. The degradation is very similar to the Gaussian case.
We conclude by tabulating the covariance estimator error for existing surveys.

\section{Simple Example}\label{toy}

Suppose the set of measurements $\data_i$ each is designed to measure a single parameter $x$, and consider the case when the covariance
matrix is diagonal, so $C_{ij}=\delta_{\ij}\sigma_i^2$. Then, the inverse of the covariance matrix $\Psi\equiv C^{-1}$ is also diagonal with elements $\Psi_i = \sigma_i^{-2}$. In this simple case, we need to minimize
\be
\chi^2(x) = \sum_i (\data_i-x)^2\Psi_i ;
\ee
in so doing, we arrive at an estimate for $x$:
\be
\hat x = \frac{\sum_i \data_i \Psi_i}{\sum_i \Psi_i}.
\ee
The uncertainty on this estimate can be obtained by computing $\langle (\hat x - x)^2\rangle$, which leads to
\be
\Delta x^2 %&=& \langle \left(S^{-1} \sum_i x_i \Psi_i - x\right)\left(S^{-1} \sum_j x_j \Psi_j - x\right)\rangle
%\vs
%&=& 
= \frac{\sum_{ij} \Psi_i\Psi_j  \langle 
 \data_i \data_j \rangle}{[\sum_i \Psi_i]^2}  - x^2.
\ee
The angular brackets around $\data_i \data_j$ refer to an average over the distribution from which the $\data_i$ are drawn. This distribution
is assumed to be Gaussian with mean $x$ and variance $C^t$, where $^t$ indicates this is the true variance, not necessarily equal to the 
covariance $C$ (or its inverse $\Psi$) used to estimate $x$.
Therefore, the variance of our estimator is
\be
\Delta x^2
%&=& S^{-2} \left[ x^2 S^2 + \sum_i C^t_{i} \Psi_i^2 \right] - 2x^2 + x^2
%\vs
= \frac{\sum_i C^t_i \Psi_i^2}{[\sum_i \Psi_i]^2}.
\ee
If we had access to the true covariance matrix, then $C^t_i \Psi_i$ would
be equal to unity and the sum in the numerator would be simply equal to that in the denominator, leaving the variance on our estimator to be $\Delta x^2 = 1/\sum_i\Psi_i$, which, in the limit
of equal errors on each of the $N_b$ measurements, reduces to the standard $\sigma^2/N_b$. 

Let's consider though the impact of not knowing exactly what the covariance matrix is. Write
\be
\Psi_i = \Psi_i^t + \Delta\Psi_i
.\ee
Then the error on $x$ is
\be
\Delta x^2 = 
\frac{1}{\left[ \sum_j (\Psi_j^t + \Delta\Psi_j)\right]^2}
\sum_i C^t_i \left[ \Psi_i^t + \Delta\Psi_i\right]^2
.\ee
Taylor expanding leads to
\be
\Delta x^2 = \frac{1}{\sum_i \Psi^t_i} + {\rm new\,terms}
.\ee
The first set of these new terms are linear in $\Delta\Psi$. These lead to fluctuations in the error, meaning that the error we assign to our estimator
will be wrong~\cite{Taylor:2012kz}. However, $\Delta\Psi$ is just as likely to fluctuate up as it is down, so the linear terms do
not lead to a systematic bias on the error, only an uncertainty on the error. The second set of terms is
quadratic in $\Delta\Psi$, and this set is more pernicious as it leads to a larger error in the estimator of $x$. 
That is, the estimated value of $x$ will be drawn from a distribution with a systematically {\it larger} variance than if the covariance matrix were known exactly. 

Let's compute this error in our simple model.
%The denominator can be written out to second order as
%\be
%\frac{1}{\left[ \sum_j (\Psi_j^t + \Delta\Psi_j)\right]^2}
%\simeq
%\frac{1}{(S^t)^2}\left[ 1 - 2\frac{\Delta S}{S^t} + 3(\frac{\Delta S}{S^t})^2  \right]
%\ee
%and the numerator as
%\be
%\sum_i C^t_i \left[ \Psi_i^t + \Delta\Psi_i\right]^2
%= S^t + 2\Delta S  + \sum_i C^t_i \Delta\Psi_i^2
%.\ee
The second order terms are
\be
\Delta x^2\Bigg\vert_{\rm second\,order}
= -  \frac{(\sum_i \Delta\Psi_i)^2}{\left[ \sum_i \Psi_i \right]^3} +  \frac{\sum_i C^t_i \Delta\Psi_i^2}{\left[ \sum_i \Psi_i \right]^2}
\eql{st}
\ee
Suppose the fluctuations in the covariance matrix are such that~\cite{Taylor:2012kz}
\be
\langle\Delta\Psi_i \Delta\Psi_j\rangle = \alpha \delta_{ij} \Psi_i^2
.\ee
Then, the first term in \ec{st} will be of order $N_b^{-2}$. The second on the other hand is of order
$N_b^{-1}$ so it dominates and we are left with
\be
\Delta x^2 = \frac{1+\alpha}{\sum_i \Psi_i } .
\ee
If the uncertainty in the covariance matrix is driven by a finite number of simulations $N_s$, then we will see that $\alpha\simeq 1/N_s$.
We call the new term {\it covariance estimator error}, and it simply increases the errors on our estimate of $x$.  Although one can drive this error down by running many simulations, the number of (expensive) simulations required in the era of percent level measurements is apparently greater than a hundred, difficult but manageable. Unfortunately, this very simple case of diagonal errors does not capture the full danger of the situation. In the more realistic case that the covariance matrix is not diagonal, $\alpha$ scales as $N_b/N_s$, so if there are measurements in a large number of bands, it will become harder and harder to reduce the covariance error.

\newcommand\xid{x^{\rm d}}

\section{Covariance Error in the General Case}       % Enter section title between curly braces
\label{sec:covariance_error_general}

We now generalize this treatment in three ways: First, we allow the covariance matrix to have off-diagonal elements, so 
$\Psi_{ij} = C^{-1}_{ij}$ is no longer just a diagonal matrix. Second, we allow for more than one parameter; instead of $x$, we envision
fitting for a full set of parameters, $p_\alpha$. Finally, the measurements are likely not direct estimates of the parameters.
If we call the data in $N_b$ bands $\xid_i$, then we want to extract values of the cosmological parameters $p_\alpha$ from these measurements. The theoretical predictions for these measurements, call them $x_i$ depend on the parameters: $x_i=x_i(p_\alpha)$, usually in some complicated way.
For simplicity, we shift all parameters so the true values are equal to 0. Then the predictions $x_i(p=0)$ are equal to the true values $\xit_i$. The measured values will not be exactly equal to $\xit$, but we expect the mean over many realizations to equal to the true set:
\be
\langle \xid_i\rangle = \xit_i
\ee
and the spread is given by the covariance matrix
\be
\ct_{ij} \equiv \langle (\xid_i-\xit_i)(\xid_j-\xit_j)\rangle.
\ee
where again superscript $^t$ denotes the true value.
We will extract the best fit values of the parameters by minimizing \ec{chisq}.
Note again that the covariance matrix here is not equal to the true one; this is the effect we want to explore: what happens to our parameter extraction when the covariance matrix is wrong?

Let's decompose the $\chi^2$ into two pieces:
\be 
\chi^2(p) = \chi_0^2(p)+ \Delta\chi^2(p)\eql{chi}
\ee
where
\be
\chi_0^2 \equiv \sum_{ij} (\xid_i-x_i(p)) \ctinv_{ij} (\xid_j-x_j(p)) 
\ee
and the term due to the uncertainty in the covariance matrix is
\be
\Delta\chi^2 \equiv \sum_{ij} (\xid_i-x_i(p)) \Delta\Psi_{ij}
 (\xid_j-x_j(p)) 
\eql{deltadef}
\ee
where
\be
\Delta\Psi_{ij} \equiv \cinv_{ij} -\ctinv_{ij}
.\ee
Both $\chi_0^2$ and $\Delta\chi^2$ are functions of $p$, and we can Taylor expand both around $p=0$. Apart from an irrelevant constant, the standard piece is
\be
\chi_0^2(p) \simeq - 2\sum_{ij} \frac{\partial x_i}{\partial p_\alpha} \ctinv_{ij} (\xid_j-\xit_j) p_\alpha+ F_{\alpha\beta} p_\alpha p_\beta\eql{chistd}
\ee
where
\bea
F_{\alpha\beta} &\equiv &\frac{1}{2} \,\frac{\partial\chi_0^2}{\partial p_\alpha\partial p_\beta} \vs
&\simeq&  \sum_{ij} \frac{\partial x_i}{\partial p_\alpha} \ctinv_{ij} \frac{\partial x_j}{\partial p_\beta} .
\eea
The approximate equality on the second line follows since operating with the derivative twice on $\xit$ leaves a factor of $\xid_i-x_i$, which
averages to zero.
Before turning to the effects of the new piece, it is worth recalling the derivation for the mean and variance of the estimator for $p_\alpha$
using the standard terms. Minimizing the Taylor expanded $\chi_0^2$ with respect to $p_\alpha$ leads to the estimator
\be
\hat p_\alpha = F^{-1}_{\alpha\beta} \sum_{ij} \frac{\partial x_i}{\partial p_\beta} \ctinv_{ij} (\xid_j-\xit_j).\eql{simest}
\ee
Since $\langle (\xid_j-x_j) \rangle=0$, the mean of this estimator is zero, equal to the true value, so the estimator is unbiased. The expected variance is obtained by squaring \ec{simest} and using the fact that $\langle(\xid_j-\xit_j)(\xid_j-\xit_j)\rangle=\ct_{jj'}$:
\bea
\langle \hat p_\alpha \hat p_{\alpha'} \rangle
&=&F^{-1}_{\alpha\beta} F^{-1}_{\alpha'\beta'}   \sum_{ij} \frac{\partial x_i}{\partial p_\beta} \ctinv_{ij} 
\frac{\partial x_{j}}{\partial p_{\beta'}} 
\vs
&=& F^{-1}_{\alpha\alpha'}
\eea
where the second equality follows from recognizing the sum over $i,j$ as the definition of $F$ and then setting $F^{-1}F=I$. So $F^{-1}$ is the projected covariance matrix on the parameters if $C$ is known exactly.

\newcommand\psit{\Psi^t}
To account for the effect of the uncertainty in the covariance matrix, we now Taylor expand $\Delta\chi^2$ in \ec{chi}:
%\sum_{ij} (\xi_i-\xit_i) \left[ \cinv_{ij} -\ctinv_{ij} \right] (\xi_j-\xit_j) 
\be
\Delta\chi^2 \simeq -2 \sum_{ij} \frac{\partial x_i}{\partial p_\alpha} \Delta\Psi_{ij} (\xid_j-\xit_j) p_\alpha + 
 \Delta F_{\alpha\beta} p_\alpha p_\beta\eql{dchi}
\ee
with
\be\label{eq:deltaF}
\Delta F_{\alpha\beta} \equiv \sum_{ij} \frac{\partial x_i}{\partial p_\alpha} \Delta\Psi_{ij} \frac{\partial x_j}{\partial p_\beta} 
.\ee
The changes to $\chi^2$ translate into a new estimator for the parameters:
\be\label{eq:phat_witherrors}
\hat p_\alpha = \left[ F+ \Delta F\right]^{-1}_{\alpha\alpha'} \frac{\partial x_i}{\partial p_{\alpha'}} \,
\left[ \psit + \Delta \Psi\right]_{ij} \left(\xid_j-\xit_j\right)
.\ee
Just as in the toy model of \S\ref{toy}, we can expand this estimator in powers of $\Delta\Psi$, and -- subject to the caveats mentioned below --
the estimator will remain unbiased but its variance will increase.

Although we are interested in the terms second order in $\Delta\Psi$ as these lead to larger errors on the parameters, it is worth pausing to comment here on two situations where the linear terms could lead to a bias: (i) when the covariance matrix depends on the parameters and this dependence is ignored by fixing $C$ and (ii) when the fluctuations in $\Delta\Psi$ are correlated with fluctuations in the data. To illustrate consider the simple situation where the elements of the inverse covariance matrix are monotonically decreasing functions of $p$ (e.g., in the diagonal case, when $p$ is the amplitude, the cosmic variance will be larger when $p$ increases and therefore elements of the inverse covariance matrix will be smaller when $p$ is greater than zero). Then, the assumed fixed value of $\Psi$ will be less than the true value when $p<0$ and greater than the true value when $p>0$; equivalently $\Delta\Psi$ will start negative and turn positive as $p$ passes through zero. If the fluctuations in $\Delta\Psi$ are uncorrelated with fluctuations in the data, then the first term in \ec{dchi} has mean zero. The second will be negative when $p<0$ and positive when $p>0$. This will then mistakenly favor regions of parameter space with $p<0$. A full understanding of the bias induced by neglecting the parameter dependence of the covariance matrix is beyond the scope of this paper (in particular, the determinant in the prefactor of the likelihood also needs to be considered)~\cite{schneider08}, but this simple example makes some of the dangers explicit.
%MDS20130328 Comments from our email exchange on 2013-03-19
%That is, models with large errors are more readily consistent with the data so the parameter-dependent
%errors must be modeled correctly to avoid a bias.
%The corollary to this that if the parameter-dependent errors are correctly modeled, then $\partial \Delta\Psi / \partial p_{\alpha}%\approx 0$ and, to lowest order, there is no bias. 
%
The second potential bias occurs when $\langle \Delta\Psi (\xid-\xit)\rangle$ is non-zero. This happens most obviously when the data itself is used to generate the covariance matrix. In that case, upwards fluctuations in the data would lead to downwards fluctuations in $\Delta\Psi$, so -- taking into account the overall minus sign -- the coefficient of the linear term in \ec{dchi} would be positive.
This change will {\it increase} the estimated value of $p$. If the fluctuation in the data was negative, there would be a positive fluctuation in $\Delta\Psi$, again leading to a positive linear coefficient in $\Delta\chi^2$. Again, the bias would push to larger values of $p$. The conclusion is that a correlation between the data and the covariance matrix may induce a parameter bias. In the simple case where the fluctuations in the covariance matrix are positive correlated with fluctuations in the data and the derivative with respect to the parameters are also monotonically increasing, the parameters will be biased high.

We now isolate terms quadratic in $\Delta\Psi$, as these lead to larger errors in the estimator:
\bea
\langle p_\alpha p_\beta \rangle\bigg\vert_{\rm s.o.}
&=&
F^{-1}_{\alpha\alpha'} \left[ 
\frac{\partial x_i}{\partial p_{\alpha'}} \,\frac{\partial  x_{i'}}{\partial p_{\beta'}} \ct_{jj'} (\Delta\Psi)_{ij}(\Delta\Psi)_{i'j'}
\right]
F^{-1}_{\beta'\beta} \vs
&&
- \left[F^{-1} \Delta F  F^{-1} \Delta F F^{-1} \right]_{\alpha\beta}\eql{pp}
.\eea
Here the angular brackets denote the expectation over the random values of $\xid$ drawn from the Gaussian distribution with mean $x(p=0)$ and
variance $\ct$. We have not (yet) computed the expectation of the fluctuations in $\Psi$.
Note that this expression reduces to \ec{st} in the 1-parameter, diagonal covariance matrix case.
To complete the calculation, we need an expression for variance of the fluctuations in $\Delta\Psi$. Let us write these generically as
\be
\langle \Delta\Psi_{ij}
\Delta\Psi_{i'j'}
\rangle
= A \Psi_{ij}\Psi_{i'j'} + B (\Psi_{ii'}\Psi_{jj'} + \Psi_{ij'}\Psi_{ji'} )
.
\label{eq:covcov_coefs}
\ee
Inserting this expression into \ec{pp} leads to
\be \label{eq:sim_cov_error}
\langle p_\alpha p_\beta \rangle\bigg\vert_{\rm s.o.}
=
B F^{-1}_{\alpha\beta} (N_b-N_p)
\ee
where $N_p$ is the number of parameters in the fit and 
has the restriction, $N_p<N_b$. Eq.~\Ec{sim_cov_error} is our main result, demonstrating that
uncertainty in the covariance matrix propagates directly to a new source of uncertainty in the 
estimate of parameters. This uncertainty is proportional to $F^{-1}_{\alpha\beta}$, which is equal to the parameter covariance
in the absence of this additional error. So covariance error does not alter the shape of the
constraints, but does inevitably lead to looser constraints.

%Eq.~(\ref{eq:sim_cov_error}) shows that the increase in the error on any 
%one parameter due to covariance estimator error 
%decreases as the number of parameters, $N_p$, increases (for fixed $N_b$). 
%However, the joint parameter constraints always increase when there 
%is an error in the data covariance estimate. Adopting the 
%definition of the joint parameter constraint 
%Figure of Merit (FoM)~\cite{wang08},
%\begin{equation}
%	{\rm FoM} \equiv \left[{\rm det}\,{\rm Cov}(p) \right]^{-1/2},
%\end{equation}
%the fractional decrease in the FoM from covariance estimator error is,
%\begin{equation}\label{eq:FoMratio}
%	\frac{{\rm FoM}}{{\rm FoM}(N_s=\infty)} = 
%	\left[1 + B(N_b-N_p)\right]^{-N_p/2}.
%\end{equation}
%In the relevant range of parameter space ($N_p\le N_b$), Eq.~(\ref{eq:FoMratio}) shows that the Figure of Merit is always smaller than the $N_s=\infty$ limit. The FoM defined in this way quantifies the extent to which the allowed region in parameter space grows in ``volume'' due to covariance estimator error.
% In practice, N_p is fixed but N_b could be varied 
% to maximize the FoM.
% There is a tradeoff between minimizing the simulation 
% covariance error and maximizing the FoM (with not cov error).

A simple way to think of this degradation is to recall that the parameter covariance matrix is inversely proportional to $f_{\rm sky}$, the fraction of sky covered by a survey. Covariance error enters in an identical way, so if the new variance captured in \ec{sim_cov_error} has coefficient $B(N_b-N_p)$ equal to 0.1, for example, the result is equivalent to throwing away 10\% of the data set.

\subsection{Gaussian limit}

Taylor et al.~\cite{Taylor:2012kz} computed the values of $A$ and $B$ in the Gaussian case (after correcting for the bias in the inverse covariance estimator~\cite{hartlap07}):
\bea \label{eq:taylorcov}
A &=& \frac{2}{(N_s-N_b-1)(N_s-N_b-4)}\vs
B &=& \frac{N_s-N_b-2}{(N_s-N_b-1)(N_s-N_b-4)}
\label{eq:BTaylor}
\eea
As in the toy model of \S\ref{toy}, in the (common) limit that $N_s\gg N_b\gg N_p$,
the variance is enhanced over the standard variance by a factor of $(1+N_b/N_s)$. 
This is our main conclusion.

\subsection{Weak Lensing Spectra}

\begin{figure}
	\centerline{
		\includegraphics[width=0.5\textwidth]{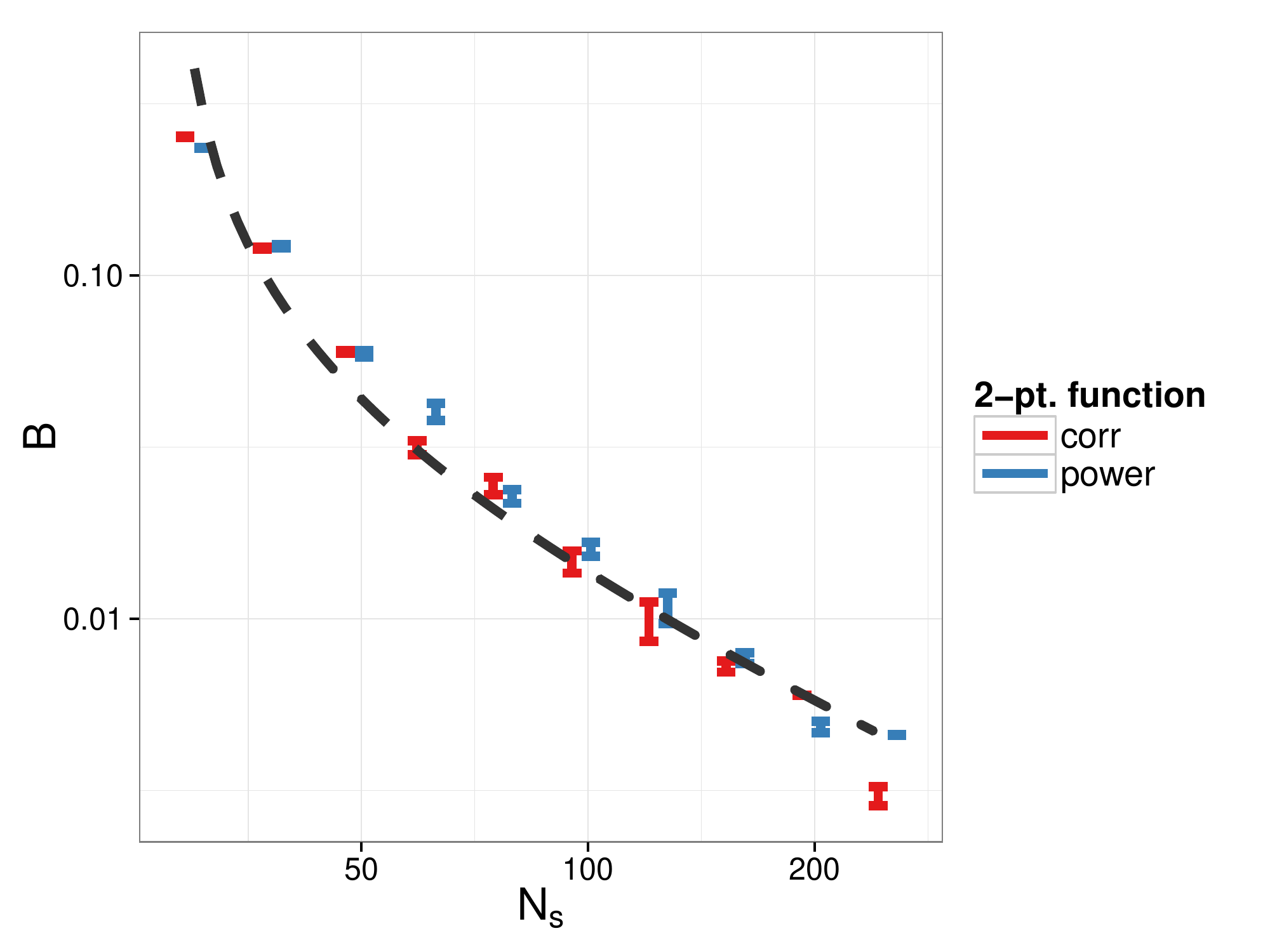}
	}
	\caption{
	The coefficient $B$ from Eq.~(\ref{eq:covcov_coefs}) as a function 
	of the number of simulation realizations using the lensing two-point correlation function (`corr')
	and power spectrum (`power') from \cite{sato09, sato11} (with a delta-function source distribution
	at $z=1$). There are 1000 realizations of the 
	simulated two-point functions. We take the sample covariance using all 1000 simulations as 
	a reference, and compare with the sample covariance using subsets of $N_s$ simulations.
	The dashed black line shows the prediction from Eq.~(\ref{eq:BTaylor}) with $N_b=24$.
	The error bars indicate the standard error on $B$ as a regression coefficient fit simultaneously 
	to the $N_b^2 \times N_b^2$ components of Cov$(\Delta\Psi)$.
	}
	\label{fig:satocomparison}
\end{figure}

We can compute covariance estimator error for non-guassian fields by using a subset of available simulations. As an example, we use the suite of weak lensing simulations from ~\cite{sato09, sato11}, assuming that the true covariance matrix is obtained from the scatter in all the simulations (1000 total). Then using only some of the simulations, we estimate $\Delta\Psi$
and therefore $B$ by taking the difference in $\Psi$ from the smaller and full set of simulations. The resulting estimate of $B$ is shown in 
Figure~\ref{fig:satocomparison} compared with the Gaussian prediction.
It is seen that, even for this highly non-gaussian field, 
Eq.~(\ref{eq:BTaylor}) gives a good fit to the 
simulation samples.
There are two reasons one might expect $B$ to exhibit a different dependence on $N_s$, 
1) the two-point function of a Gaussian random field is not itself Gaussian distributed,
2) nonlinear gravitational evolution skews the statistics of the cosmological mass 
density field away from Gaussian.
However, because the two-point function estimator is a sum of squares of the density perturbations, 
the distribution of the {\it estimator} may tend to a Gaussian as the number of modes 
in a (wavenumber or angular) bin becomes large. Figure~\ref{fig:satocomparison} is consistent 
with this explanation.

%%%%%%%%%%%%%%%%%%%%%%%%%%%%%%%%%%%%%%%
\subsection{Current surveys}

Table~\ref{tab:coverrors} demonstrates the effect of simulation 
covariance error for some recently published cosmological surveys (which estimated covariance matrices from simulation realizations rather than from the data). We find that the degradation ranges from 5-15\%.
%Figures of Merit are smaller than they would be with an infinite number of simulations by factor of order 10\% for DLS up to a factor of 3 for CHFTLens. The true degradation is probably not as severe as suggested in Table~\ref{tab:coverrors} since prior constraints (e.g. from the CMB) were used, but the Table does 
%indicate that covariance estimator error needs to be taken seriously even for current surveys.
\begin{table}[ht]
\begin{center}
\caption{\label{tab:coverrors}Increase in the variance of each parameter due to covariance estimator errors for some recently published
survey analyses.}
\begin{tabular}{lcccc}
\hline\hline
Survey & $N_s$ & $N_b$ && Fractional Increase \\
& & && in Variance \\
\hline 
BOSS~\cite{sanchez12}       & 600  & 41  && 7\%\\
DLS~\cite{jee13}            & 1000 & 60   && 6\% \\
CHFTLens~\cite{kilbinger13} & 184  & 24&  &  13\%  \\
\hline\hline 
\end{tabular}
\end{center}
\end{table}

\section{Conclusions}

We derived a new contribution to parameter uncertainties from the uncertainty in sample 
data covariance matrices estimated from simulations. This error adds in quadrature with other 
sources of parameter uncertainty and scales with the ratio of the number of data bins to the 
number of simulation realizations.

Current surveys use hundreds of simulations, but even this large number leads to an underestimate of parameter uncertainties 
by $\sim$5-15\%. 
Future surveys, which will be sensitive enough to measure in hundreds of bins will require of order $10^4$ simulation realizations 
(per cosmological model) to 
prevent 5-10\% degradation in the parameter uncertainties. 
Mitigation schemes such as shrinkage estimators~\cite{pope08}, emulators~\cite{schneider08, morrison13},
and large-scale mode-resampling~\cite{schneider11} will be important to 
reduce these computational requirements to tractable levels.

When the covariance matrix varies with cosmology (as is generally the case), 
there will be additional contributions to the covariance estimator error. 
We will derive these contributions in future work, but expect them to 
be sub-dominant to the primary result we present in this paper as long as the model for the cosmology-dependent covariance is accurate enough to ensure that the fluctuations, $\Delta\Psi$, in the covariance estimator are approximately independent of cosmology.

\acknowledgments{SD is supported by the U.S.
Department of Energy, including grant DE-FG02-95ER40896. Part of this work was performed under the auspices of the U.S. Department of Energy by Lawrence Livermore National Laboratory under Contract DE-AC52-07NA27344.}

\bibliography{covariance}

% Set the ending of a LaTeX document
\end{document}